\documentclass{iaus}
\usepackage{graphicx}

\title[VMC \& the SFH of Local Galaxies ] %% give here short title %%
{The VMC survey and the SFH of some Local Group Galaxies}

\author[Maria-Rosa L. Cioni]   %% give here short author list %%
{Maria-Rosa L. Cioni$^1$}
\affiliation{$^1$SUPA, School of Physics, University of Edinburgh,
Edinburgh, EH9 3HJ, UK \break email: mrc@roe.ac.uk\\[\affilskip]}

\pubyear{2007}
\volume{241}  
\pagerange{}
\date{?? and in revised form ??}
\jname{Stellar Populations as Building Blocks of Galaxies}
\editors{ }
\begin{document}

\maketitle

\begin{abstract}
  
  The $K_s$-band magnitude distribution of carbon-rich and oxygen-rich
  asymptotic giant branch stars within Local Group galaxies like the
  Magellanic Clouds, NGC 6822, M33 and SagDIG is easily obtained from
  ground-based observations. Appropriate stellar evolutionary models
  covering a range of metallicities and star formation rates are used
  to produce theoretical distributions that allow us to derive the
  history of star formation across these galaxies. I will show the
  result of these studies and discuss the application of this
  technique to more distant systems as well as deeper observations,
  like those that VISTA will provide, to improve our understanding of,
  in particular, the Magellanic Clouds.


\end{abstract}

\firstsection % if your document starts with a section,

\section{The VMC survey}

\begin{figure}
  \includegraphics[height=6cm,width=13.5cm]{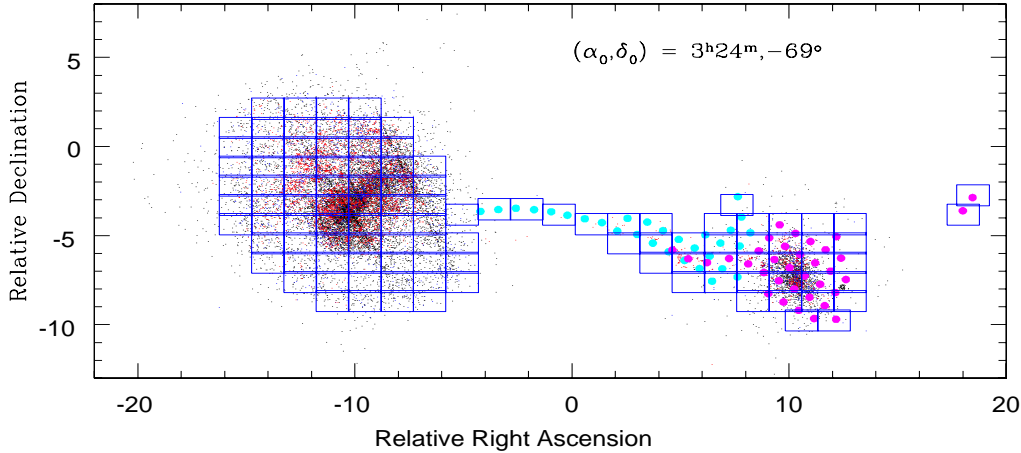}
  \caption{Distribution of VISTA tiles across the Magellanic
  System. Underlying small dots indicate the distribution of C stars,
  clusters and associations while thick dots show the location of VST
  pointings.}
\label{vmc}
\end{figure}

The VISTA Public Survey of the Magellanic System (VMC) has been
recommended by the ESO Public Survey Panel and the ESO 
Observing Programme Committee to form part of the core programme of
VISTA. This new telsecope, soon to be commissioned, will host a
near-infrared wide-field camera (VIRCAM). The VMC originated from an
international collaboration and will provide the community with a
unique dataset to study the Magellanic System stellar population and
structure. 

These data will cover entirely the major components of the system
(Fig.~\ref{vmc}): a wide area comprising the Large Magellanic Cloud (LMC), the
Small Magellanic Cloud (SMC), the Bridge connecting them and a few
fields in the Stream where the stellar content is expected to be
large. Across five years observations in the $Y$, $J$ and $K_s$
filters will be obtained. Three epochs in $Y$ and $J$ and $12$ epochs
in $K_s$ will constrain the mean magnitude of short-period variable
(RR Lyrae and Cepheid stars) and the periodicity of long period
variable stars. Both the Bridge and SMC areas will also be covered at
optical wavelengths by the VLT Survey Telescope (VST).

The main VMC science goals are: to derive the global and spatially
resolved star formation history (SFH) of the system and its
three-dimensional (3D) geometry. These require a depth of $K_s=20.3$
with a S/N$=10$. The SFH will be measured from the interpretation of
mainly giant stars and main sequence turn off stars using stellar
evolution models as well as three-body simulations of the system. The
3D-geometry will be obtained from: the period-magnitude relation of RR
Lyrae and Cepheid stars (where the period will come from large-scale
optical surveys like EROS-II for the LMC and VST observations for the
other components of the system), the distribution of stars in the red
clump phase of evolution and stellar clusters. VMC data will also
allow us to reduce the uncertainty on the measurement of the distance
to the LMC by a factor of two, to derive the mass of obscured AGB
stars discovered recently by the Spitzer infrared space telescope, to
study the formation of stars of about a solar mass, to uncover the
missing number of Planetary Nebulae and eventually to constrain the
proper motion of the Magellanic Clouds that the ESA space mission GAIA
will accurately measure.

\section{Metallicity and SFR of Local Galaxies}

A near-infrared wide-field view of galaxies allow us to derive their
global properties and, in particular, to study the distribution of the
mean-age and metallicity of their stellar population. This work, using
data from the DENIS and 2MASS surveys has been published this year on
the Magellanic Clouds (\cite[Cioni, \etal\ 2006a, 2006b, 2006c]{}). A
similar work on the Magellanic type galaxy NGC 6822 has recently been
submitted (\cite[Cioni, \etal\ 2007a]{cioni07a}), as well as on the
faint irregular galaxy SagDIG (\cite[Gullieuszik, \etal\ 
2007]{marco07}), while a study of the nearby spiral galaxy M33 is
currently in preparation (\cite[Cioni, \etal\ 2007b]{cioni07b}).

\subsection{The $K_s$ method}

Carbon-rich (C-rich) and oxygen-rich (O-rich) asymptotic giant branch
(AGB) stars can be distinguished using the near-infrared
colour-magnitude diagram ($J-K_s$, $K_s$). Other criteria involve
optical broad- or narrow-band photometry or low-resolution
spectroscopy. Their observed number distribution as a function of
$K_s$ magnitude compared with theoretical distributions created using
theoretical models allow us to create maps of the most probable
metallicity (Z) and star formation rate (SFR) of the underlying
stellar population.  The comparison, evaluated using the $\chi ^2$
test, has been made across spatial regions (sectors of elliptical
coronae) containing a statistical significant number of stars. For
example, the subdivision of the area covered by M33 comprises $3500$
O-rich AGB stars and $300$ C-rich AGB stars in each of five concentric
ellipses; a histogram was created for each of eight sectors per ellipse.

Theoretical distributions have been created using the TRILEGAL code to
simulate stars according to a SFR, age-metallicity relation and
initial mass function. In particular, luminosity, effective
temperature and gravity have been interpolated among stellar
evolutionary tracks from: \cite[Bertelli, \etal\ (1994)]{bertelli94}
for massive stars and \cite[Girardi, \etal\ (2000)]{girardi00} for low-
and intermediate-mass stars including the recipe for thermal pulsing
AGB stars by \cite[Marigo, Girardi \& Bressan (1999)]{marigo99}.
Bolometric tables were used to derive magnitudes and include
photometric errors.  A combination of different model parameters (SFRs
and Zs) results in a different number of C- and O-rich AGB stars as
well as in a different location in the ($J-K_s$, $K_s$) space. The
best fit model ($=$ the lowest $\chi ^2$) is assigned to a given area
and a spatial map of best Z and best SFR is created. 

\begin{figure}
 \includegraphics[height=9cm,width=12cm]{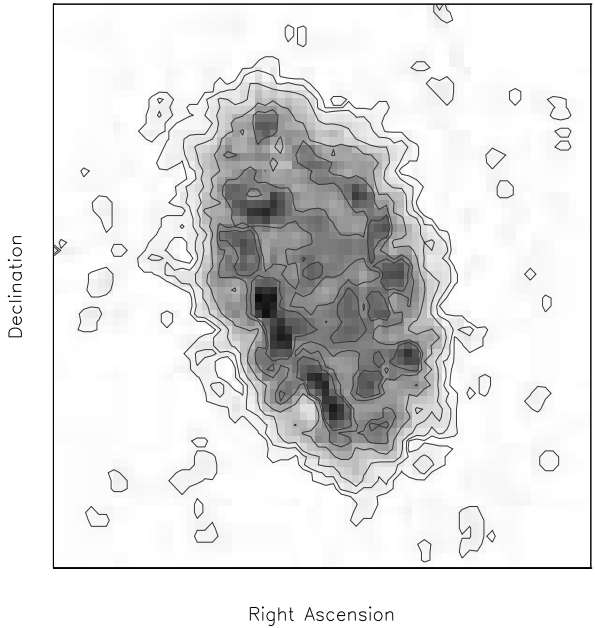}
 \caption{Distribution of the C/M ratio across M33. C-rich and O-rich AGB stars above the tip of the RGB have been selected using the colour-magnitude diagram ($J-K_s$, $K_s$). Darker regions correspond to higher ratios. Contours are at: $0.25$, $0.5$, $1.0$, $2.0$, $3.0$, $3.5$ and $4.0$.}
\end{figure}

\subsection{Applications: main results}

The LMC stellar population appears younger in the East than in the
West and the bar has a composite stellar population, see also 
\cite{cioni07c}. Metallicity maps do agree with the distribution
shown by the C/M ratio (the ratio between C-rich and O-rich AGB
stars). Maps were also corrected for the orientation of the galaxy in
the sky and for the effect, if present, of differential extinction.

The SMC stellar population appears more metal rich in a ring
surrounding the central part of the galaxy. A region of increasing
metallicity moves anti-clockwise with increasing time perhaps tracing
the dynamical evolution of the galaxy and/or the propagation of star
formation in agreement with the results suggested by \cite[Harris \&
Zaritsky (2004)]{harris04}. 

The stellar population of NGC 6822 derived from observations obtained
at the WHT in La Palma appears on average $8$ Gyr old. There is
excellent agreement between the overall metallicity trend with
direct measurements of [Fe/H] in individual stars in the inner
galaxy. The outer halo appears metal poorer and older.

The stellar population of M33 derived from observations obtained at
the UKIRT telescope appears on average $7-8$ Gyr old and it is metal
poor.  The distribution of high C/M ratio values, corresponding to a
lower metallicity (\cite[Battinelli \& Demers 2005]{battinelli05}),
traces the structure of the major spiral arms presenting a puzzle to
its interpretation. Usually, spiral arms are associated with the
youngest and therefore the metal richest regions of a galaxy. The
effect of differential reddening and of the structure of the galaxy on
the C/M ratio is negligible while it is fundamental in the
distribution of metallicity and SFR obtained from the number density
of AGB stars.  This correction has not yet been implemented and, in
fact, preliminary results suggest that the North-East of the galaxy is
more obscured than the South-West of it.

Although SagDIG contains a limited population of intermediate-age
stars, C-rich AGB stars were carefully identified and their
distribution analyzed as for the previous galaxies. Near-infrared
observations were, in this case, obtained from the NTT telescope in La
Silla. The population of SagDIG appears on average $4$
Gyr old and with [Fe/H]$< -1.3$ dex.

\section{Conclusions}

Inhomogeneities in Z and SFR as those derived here can be interpreted
via simulations of the spatial distribution of stars.  This has been
done for the LMC accounting also for the effect of its interaction
with the SMC and the Milky Way, see \cite{bekki07}. If stars formed in
clumps of $\le 10^7 M_\odot$ or smaller they constitute today the
field stellar population of the LMC. Each clump has a certain age and
metallicity and although stars from different clumps appear spatially
mixed, it is possible that they do preserve the age and metallicity of
the gas cloud from which they formed. Thus, identifying these
components provides a fossil record of the galaxy history which
ultimately will be explained with the availability of the kinematics
information on the different stellar populations.

The $K_s$ method will be applied to other galaxies in the Local Group
as soon as suitable data will become available, but also to more
distant galaxies resolved into stars.
Understanding giant stars is a key to understand not only an extreme
aspect of stellar evolution, but also to understand global properties
of galaxies and to trace their history. These studies are essential
for new sensitive instruments which probe the Universe at
near-infrared wavelengths!

\begin{acknowledgments}
  I would like to acknowledge the collaborative work of Leo Girardi,
  Paola Marigo and Harm Habing, especially the theoretical effort, on
  the reconstruction of the metallicity and mean-age distribution of
  galaxies. I would also like to thank the VMC team for contributing
  to the success of the VMC proposal.
\end{acknowledgments}

\end{document}